\documentstyle[aps,prb,epsf,multicol,amsmath,graphics]{revtex}

\begin{document}
\draft

\title{
  Charge excitations in NaV$_{2}$O$_{5}$
}

\author{A. H\"{u}bsch, C. Waidacher, and K. W. Becker}
\address{
  Institut f\"{u}r Theoretische Physik,
  Technische Universit\"{a}t Dresden, D-01062 Dresden, Germany 
}

\author{W. von der Linden}
\address{
  Institut f\"{u}r Theoretische Physik, Technische Universit\"{a}t Graz,
  Petersgasse 16, A-8010 Graz, Austria
}

\date{\today}
\maketitle

\begin{abstract}
  We calculate the electron-energy loss spectrum and the optical conductivity
  for NaV$_{2}$O$_{5}$ using the standard Lanczos algorithm. The vanadium 
  ions in NaV$_{2}$O$_{5}$ form a system of coupled ladders which can be 
  described by a quarter-filled extended Hubbard model. Since this system has 
  a large unit cell, one has to be very careful to avoid finite-size effects 
  in the calculations. We show this by performing exact diagonalization of 
  different clusters with up to 16 sites. The calculated loss function for the 
  extended Hubbard model shows good agreement with experimental 
  spectra. Furthermore, a qualitative description of the optical conductivity 
  is obtained with the same Hamiltonian, and the same set of model parameters. 
  The comparison with the experiment shows that interladder hopping is of 
  minor importance for a realistic description of charge excitations in 
  NaV$_{2}$O$_{5}$. We find that the character of the excitations depends 
  strongly on the direction of momentum transfer.
\end{abstract}


\widetext
\begin{multicols}{2}
\narrowtext


\section{Introduction}

The insulating system $\alpha'$-NaV$_{2}$O$_{5}$ belongs to the fascinating 
class of highly correlated low-dimensional electronic systems. Recently, its 
physical properties have been intensively investigated theoretically as well 
as experimentally. The magnetic susceptibility of NaV$_{2}$O$_{5}$, can be 
well described by a $S=1/2$ antiferromagnetic Heisenberg chain with exchange 
interactions of $J=440$K and $560$K for temperatures below\cite{Weiden} and 
above\cite{Isobe} the transition temperature $T_{C}\approx 34$K. The opening 
of a spin gap at the phase transition is accompanied by unit cell doubling in 
the $a$ and $b$ direction, and quadrupling in the $c$ direction.\cite{Fujii} 
Based on an early X-ray study\cite{Capry} which postulated two different V 
sites, i.e. magnetic V$^{4+}$ chains along the $b$ axis separated by 
non-magnetic V$^{5+}$ chains, $\alpha'$-NaV$_{2}$O$_{5}$ was initially 
identified as an inorganic spin-Peierls (SP) material\cite{Isobe} similar to 
CuGeO$_{3}$. However, according to a recent crystal structure 
analysis,\cite{Schnering} at room temperature all V sites are equivalent with 
a formal valence $+4.5$. Therefore, the V sites form a quarter-filled 
ladder\cite{Gros} which can be mapped onto a $S=1/2$ chain,\cite{Horsch} and 
the resulting exchange couplings $J$ along the chain agree well with the 
experimental data. In principle, this effective chain may show all features of 
an ordinary one-dimensional $S=1/2$ Heisenberg chain, including the 
spin-Peierls transition. On the other hand, various experimental results for 
NaV$_{2}$O$_{5}$ indicate a more complicated transition at $T_{C}$: The BCS 
ratio of $2\Delta/kT_{C}$ for the dimerization gap $\Delta$ at zero 
temperature has the anomalously high value of $6.44$. The phase transition 
consists of two very close transitions,\cite{Koeppen} and two inequivalent V 
sites were detected by NMR,\cite{Ohama} and attributed to V$^{4+}$ and 
V$^{5+}$. This has motivated both theoretical and experimental work, proposing 
a pure charge ordering (CO) instability\cite{Seo,Mostovoy} or CO coupled to SP 
distortion.\cite{Thalmeier,Fagot} Recently, the space group $Fmm2$ for the 
low-temperature crystal structure was determined from x-ray scattering 
data,\cite{Luedecke} and three inequivalent V sites in the V$^{4+}$, V$^{5+}$, 
and V$^{4.5+}$ oxidation states were proposed.\cite{Smaalen,Boer} Since this 
space group is incompatible with the NMR measurement the low-temperature 
crystal structure and the charge ordering pattern are still under 
discussion.\cite{Ohama2,Bernert}

Charge ordering can be viewed as a Wigner-crystallization on a 
lattice,\cite{Fulde} which is caused by long-range Coulomb interaction. Thus 
different models with nearest-neighbor Coulomb repulsions were 
studied\cite{Seo,Vojta1,Vojta2} and an in-line\cite{Thalmeier} and a zig-zag 
structure\cite{Seo,Mostovoy} for NaV$_{2}$O$_{5}$ were proposed. The 
calculation of excitation spectra is a powerful test for these models, as they 
can be directly compared with experimental data. 

Recently, we studied\cite{Atzkern} the dynamic dielectric response of 
NaV$_{2}$O$_{5}$ using an effective quarter-filled ladder model (see 
Sec.~\ref{general}). In the present paper we improve on these calculations by 
treating the complete extended Hubbard model. Furthermore, we study the role 
of the interladder hopping in more detail. We analyze possible finite-size 
effects in cluster calculations for NaV$_{2}$O$_{5}$, and in particular, we 
discuss the optical conductivity. While in previous 
work\cite{Nishimoto1,Nishimoto2} different models were used for the 
calculation of the optical conductivity and the dynamical density correlation 
function, we describe both with the same Hamiltonian, and the same set of 
model parameters.

The paper is organized as follows. In Sec.~\ref{general}, general aspects 
of our approach are presented. The value of the interladder hopping is 
discussed in Sec.~\ref{hopping}. Finite-size effects of cluster calculations 
are investigated in Sec.~\ref{finite}. Section \ref{Elect} contains the 
comparison of our results of the dynamic dielectric response with experimental 
spectra of NaV$_{2}$O$_{5}$, and in Sec.~\ref{optic_cond} we discuss the 
optical conductivity. Finally, the conclusions are presented in 
Sec.~\ref{conclusion}.


\section{General Aspects of the calculations}\label{general}

\begin{figure}
  \begin{center}
    \scalebox{0.7}{\includegraphics[180,280][422,610]{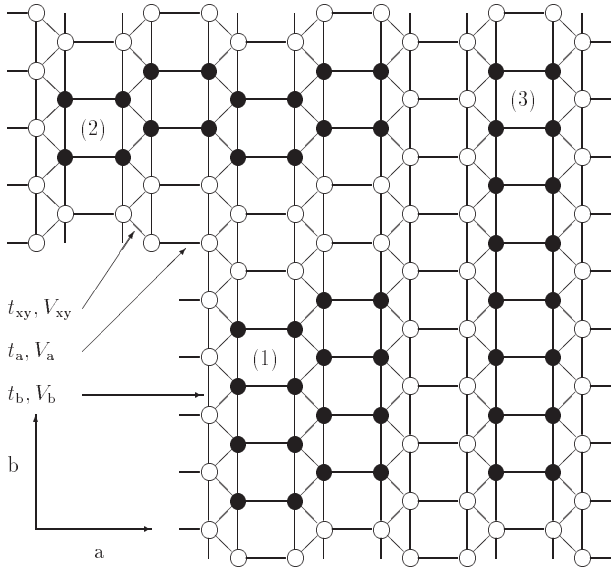}}
  \end{center}
  \caption{
    Schematic structure of the (a,b) planes in ${\rm NaV_{2}O_{5}}$ where
    the circles denote the vanadium sites. The black sites define clusters
    used in the calculation.
  }
  \label{ebene}
\end{figure}

We study the dynamic dielectric response and the optical conductivity assuming 
that the electrons in NaV$_{2}$O$_{5}$ can be described by a quarter-filled 
extended Hubbard model
\begin{eqnarray}
  {\cal H} &=& -\sum_{\langle i,j\rangle,\sigma}t_{ij}
               \left(
                 c_{i,\sigma}^{\dagger}c_{j,\sigma}+{\rm H.c.}
               \right)
               + U\sum_{i}n_{i\uparrow}n_{i\downarrow}+\nonumber\\
            && +\sum_{\langle i,j\rangle}V_{i,j}n_{i}n_{j} \label{hamilton}
\end{eqnarray}
for the 2D system shown in Fig.~\ref{ebene}. $\langle i,j\rangle$ denotes 
summation over all pairs of nearest neighbors, and spin 
$\sigma=\uparrow,\downarrow$. $c_{i,\sigma}^{\dagger}$ are electron 
creation operators, $n_{i}=\sum_{\sigma}c_{i,\sigma}^{\dagger}c_{i,\sigma}$ is 
the occupation-number operator, and $U$ denotes the Coulomb repulsion between 
electrons on the same site.  The hopping parameters $t_{ij}$, and intersite 
Coulomb interactions $V_{ij}$ are defined in Fig.~\ref{ebene}. Using 
second-order perturbation theory, one can transform Hamiltonian 
\eqref{hamilton} into an effective $t$-$J$-$V$ model
\begin{eqnarray}
  {\cal H} &=& -\sum_{\langle i,j\rangle,\sigma}t_{ij}
               \left(
                 \hat{c}_{i,\sigma}^{\dagger}\hat{c}_{j,\sigma}+{\rm h.c.}
               \right) +\nonumber\\
           &&  + \sum_{\langle i,j\rangle}J_{i,j}
               \left(
                 {\rm \bf S}_{i}\cdot{\rm \bf S}_{j}-\frac{1}{4}n_{i}n_{j}
               \right)
               +\sum_{\langle i,j\rangle}V_{i,j}n_{i}n_{j}, \label{ef_hamilton}
\end{eqnarray}
where $\hat{c}_{i,\sigma}^{\dagger}=c_{i,\sigma}^{\dagger}(1-n_{i,-\sigma})$ 
are constrained electron creation operators, and ${\rm \bf S}_{i}$ denotes the 
spin-$\frac{1}{2}$ operator at site $i$. The exchange interactions between 
vanadium neighbors are parameterized as $J_{ij}=4t_{ij}^{2}/U$.

The loss function in EELS experiments is directly proportional to the dynamic 
density-density correlation function.\cite{Schnatterly} By including the 
long-range Coulomb interaction in the model within a random-phase 
approximation (RPA) one finds for the loss function 
\begin{eqnarray}
  L(\omega,{\bf q}) & = & 
    {\rm Im}\left[
      \frac{-1}{1+v_{\bf q}\chi_{\rho}^{0}(\omega,{\bf q})}
    \right],\label{lost}
\end{eqnarray}
where
\begin{eqnarray}
  \chi_{\rho}^{0}(\omega,{\bf q}) & = &
    \frac{i}{\hbar}\int_{0}^{\infty}dt\,e^{i\omega t}\langle 
      0|[ \rho_{\bf q}(t), \rho_{-{\bf q}}]|0
    \rangle\label{response}
\end{eqnarray}
is the response function at zero temperature for the short-range interaction 
models \eqref{hamilton} and \eqref{ef_hamilton}. $\chi_{\rho}^{0}$ depends on 
the energy loss $\omega$ and the momentum transfer ${\bf q}$. $|0\rangle$ is 
the ground state, $\rho_{\bf q}$ denotes the Fourier transform of $n_{i}$, and 
$v_{\bf q}=e^{2}N/(\epsilon_{0}\epsilon_{\rm r}v{\bf q}^{2})$ is the 
long-range Coulomb interaction with unit cell volume $v$. $N$ is the number of 
electrons per unit cell, and $\epsilon_{o}$ is the permittivity. The real part 
$\epsilon_{\rm r}$ of the dielectric function can be obtained from the 
experiment. In the case of NaV$_{2}$O$_{5}$, one finds for momentum transfer 
in $a$ direction $\epsilon_{\rm r}=7$, and in $b$ direction 
$\epsilon_{\rm r}=5$.\cite{Atzkern} 

The response function $\chi_{\rho}^{0}(\omega,{\bf q})$ and the optical 
conductivity $\sigma_{\alpha}(\omega)$ are connected by the Kubo-Green 
relation
\begin{eqnarray}
  \sigma_{\alpha}(\omega) &=& 
    \epsilon_{0}\omega\lim_{{\bf q}\rightarrow 0}{\rm Im}\left[v_{\bf q}
    \chi_{\rho}^{0}(\omega,{\bf q})\right],\qquad 
    {\bf q}\parallel\alpha,\label{connect}
\end{eqnarray}
where at zero temperature $\sigma_{\alpha}(\omega)$ is defined by
\begin{eqnarray}
  \sigma_{\alpha}(\omega) &=& \frac{1}{\omega}{\rm Re}\int_{0}^{\infty}dt\, 
    e^{i\omega t}\langle 0| j_{\alpha}(t)j_{\alpha}|0\rangle.\label{optic}
\end{eqnarray}
Here $j_{\alpha}$ with $\alpha=a,b$ are the components of the current 
operator parallel to $a$ or $b$ direction. 

Equations \eqref{response}, \eqref{optic} for the response function and the 
optical conductivity  are valid for zero temperature, whereas the experiments 
have been carried out at finite temperatures.\cite{Atzkern} However, for 
NaV$_{2}$O$_5$ one finds both experimentally\cite{Atzkern,Long,Presura} and 
theoretically\cite{Cuoco} that the spectra depend only weakly on temperature. 
Therefore, we may restrict ourselves to zero temperature. Equations 
\eqref{lost} and \eqref{optic} are evaluated by direct diagonalization using 
the standard Lanczos algorithm\cite{Lin} which is limited to small clusters. 
For that reason, one can only observe localized excitations, and the 
calculation of the loss function \eqref{lost} is limited to momentum transfers 
${\bf q}\geq 2\pi/L$, where $L$ is the cluster size in ${\bf q}$ direction.


\section{Large or small interladder hopping}\label{hopping}

LDA band structure calculations\cite{Gros} and estimations based on empirical 
rules\cite{Horsch} lead to similar values of the hopping amplitudes $t_a$, 
$t_b$ perpendicular to and along the ladders. However, there are significantly 
different estimations of the interladder hopping $t_{xy}$. A small value of 
$t_{xy}=0.012$eV is found in LDA calculations.\cite{Gros} Additional evidence 
for a small $t_{xy}$ follows from the weak magnetic dispersion along the 
$a$-axis as observed by neutron scattering.\cite{Yosihama} On the other hand, 
a much larger value of $t_{xy}=0.3$eV was found from an estimation in 
Ref.~\ref{Horsch}. A relatively large interladder hopping of $t_{xy}=0.15$eV
was also obtained from a comparison of calculated and experimental optical 
conductivity.\cite{Cuoco} 

In Ref.~\onlinecite{Cuoco}, a finite temperature Lanczos method\cite{Jaklic} 
has been used to calculate the optical 
conductivity. The Lanczos approach is limited to small systems, and a cluster 
consisting of two ladders with four rungs [cluster (1) in Fig.~\ref{ebene}] 
has been used in Ref.~\onlinecite{Cuoco}. However, when a calculation is 
restricted to finite systems, finite-size effects may distort the results. 
Finite-size effects affect delocalized excitations, which are important if the 
hopping parameters $t_{ij}$ are large, and/or the momentum transfer 
${\bf q}$ is small. For that reason, the loss function $L(\omega,{\bf q})$ 
for small ${\bf q}$ is sensitive to finite-size effects. The calculation of 
the optical conductivity $\sigma_{\alpha}(\omega)$ is even more problematic, as
it is obtained from $\chi_{\rho}^{0}(\omega,{\bf q})$ in the limit 
${\bf q}\rightarrow0$ [see Eq.~\eqref{connect}]. Since the importance of 
delocalized excitations can be decreased by increasing ${\bf q}$,  we first 
study finite-size effects of the loss function $L(\omega,{\bf q})$ 
(Sec.~\ref{finite}). Already at finite ${\bf q}$ we observe large effects 
even when a rather small value of the interladder hopping $t_{xy}$ is used. 
This also means that the optical conductivity for the same cluster with a 
larger value of $t_{xy}$ is even more affected by finite-size effects (see 
Sec.~\ref{optic}). Consequently, we can show that the results for the optical 
conductivity from Ref.~\onlinecite{Cuoco} are not converged and cannot be used 
to support a large value of $t_{xy}$.

In the following, we shall use the hopping parameters 
($t_{a}=0.38$eV, $t_{b}=0.17$eV, $t_{xy}=0.012$eV) and the on-site Hubbard 
interaction ($U=2.8$eV) from Ref. \ref{Gros}. Up to now the values of the 
intersite Coulomb interactions $V_{a}$, $V_{b}$ and $V_{xy}$ are not known 
exactly. Therefore we choose moderate values for $V_{a}$ and 
$V_{b}$, so that the system is close to the quantum critical point caused by 
charge ordering.\cite{Vojta1,Vojta2,Cuoco}


\section{Finite-size effects}\label{finite}

To investigate finite-size effects, one has to use clusters of different size. 
However, the cluster consisting of two ladders with four rungs [(1) in 
Fig.~\ref{ebene}] is the largest to which the standard Lanczos 
algorithm can be applied at present. One way to overcome this problem is to 
enlarge the cluster in the direction of the momentum transfer, and to reduce 
it in perpendicular direction (so that the number of sites is kept constant). 
Of course, first one has to check that one does not distort the results by 
reducing the cluster perpendicular to the momentum transfer. To show this we 
use the effective $t$-$J$-$V$ model \eqref{ef_hamilton} with the above values 
of $t_{ij}$ and $U$ from Ref.~\ref{Gros}. The intersite Coulomb interactions 
are chosen $V_{a}=0.8$eV, $V_{b}=0.6$eV, and $V_{xy}=0.9$eV. First we compare 
the loss function of the original cluster [indicated by (1) in 
Fig.~\ref{ebene}, with periodic boundary conditions] with those of two smaller 
clusters (not shown). The first consists of two ladders with two rungs for 
momentum transfer ${\bf q}$ parallel to $a$ direction. The second cluster is 
an isolated ladder with four rungs for ${\bf q}$ parallel to $b$ direction. 
Note that periodic boundary conditions for these two smaller clusters make 
only sense in direction of the momentum transfer. Therefore we choose open 
boundary conditions perpendicular to the momentum transfer. Moreover, for the 
first of the two smaller clusters (momentum transfer in $a$ direction) one has 
to use renormalized intersite Coulomb interactions $\bar{V}_{a}=V_{a}+V_{b}$ 
and $\bar{V}_{b}=2V_{b}$. This follows from a straightforward analysis of the 
influence of adjacent rungs on the same ladder.

\begin{figure}
  \begin{center}
    \scalebox{0.59}{
      \includegraphics*[105,580][510,790]{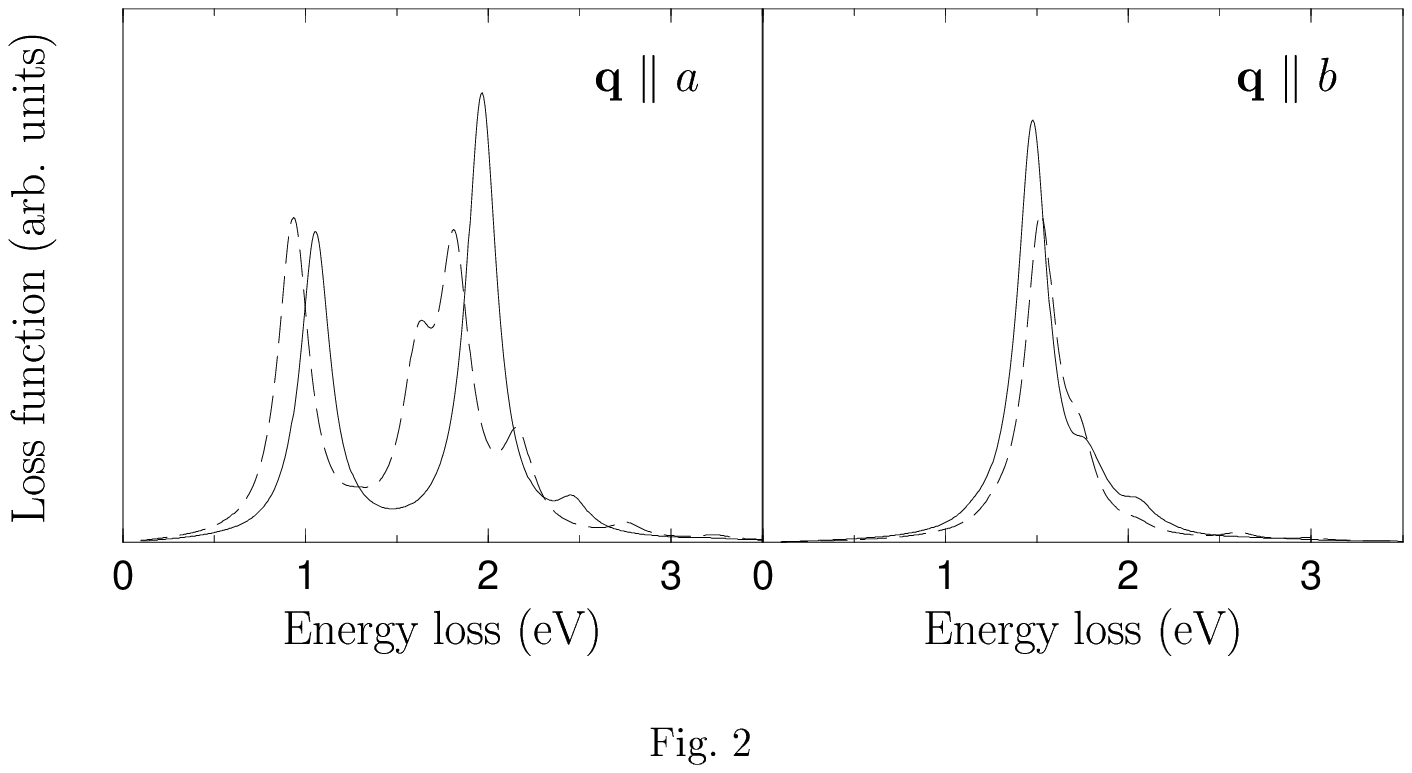}
    }
  \end{center}
  \caption{
    Comparison of the loss function for two smaller clusters (full lines, see
    the text) and the system consisting of two ladders with four rungs [dashed
    lines, cluster (1) in Fig.~\ref{ebene}]. We use the hopping parameters and
    the on-site Hubbard interaction of the $t$-$J$-$V$ model from
    Ref.~\ref{Gros}, the intersite Coulomb interactions are $V_{a}=0.8$eV,
    $V_{b}=0.6$eV, and $V_{xy}=0.9$eV. The momentum transfer
    ${\bf q}=0.3$\AA$^{-1}$ is parallel to $a$ direction (left panel) or $b$
    direction (right panel). The theoretical line spectra are broadened with
    Gaussian function of width $0.1$eV.
  }
  \label{rech_1}
\end{figure}

The results for the different clusters are shown in Fig.~\ref{rech_1}. The loss
function for the original cluster (1) of Fig.~\ref{ebene} is drawn with dashed 
lines. The results for the smaller clusters with ${\bf q}$ in $a$ or $b$  
direction are shown as solid lines in Fig.~\ref{rech_1} (left or right panel, 
respectively). As can be seen from Fig.~\ref{rech_1}, apart from a small 
splitting  of the second peak in the spectrum for {\bf q} parallel to $a$ 
direction there is good agreement between the results for the large cluster 
and both smaller systems. (Note that we use a small interladder hopping 
$t_{xy}$.)

\begin{figure}
  \begin{center}
    \scalebox{0.59}{
      \includegraphics*[105,420][510,790]{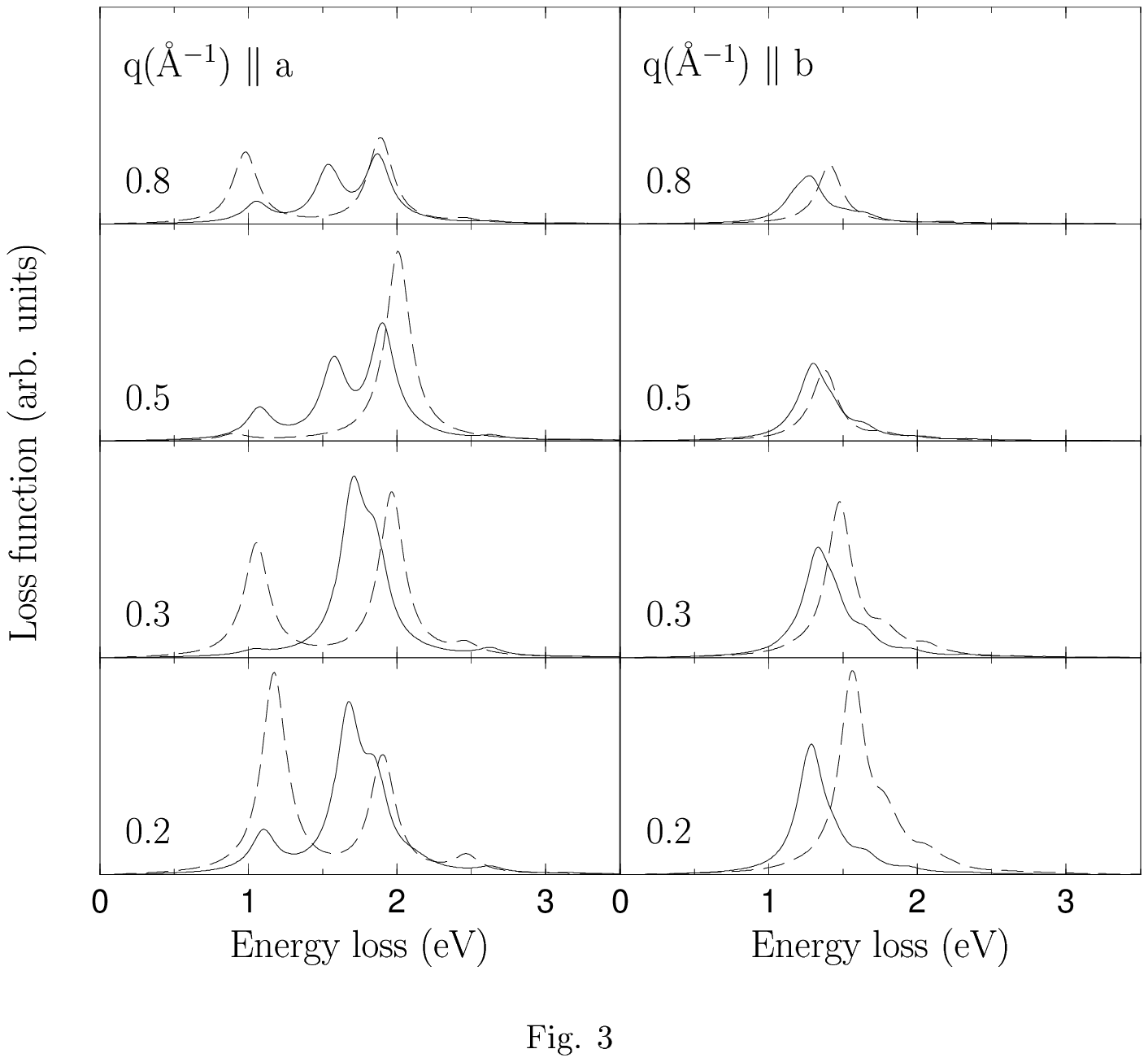}
    }
  \end{center}
  \caption{
    Finite-size effects of the loss function for clusters with eight
    rungs (full lines) and four rungs (dashed lines); parameters as for
    Fig.~\ref{rech_1}.
  }
  \label{rech_2}
\end{figure}

Now we can study finite-size effects by increasing the two smaller clusters in 
the direction of momentum transfer. In this way we obtain clusters (2) and (3) 
in Fig.~\ref{ebene}. In the left panel of Fig.~\ref{rech_2} the loss function 
for cluster (2) with four ladders (and for the previous cluster with two 
ladders) are shown for different ${\bf q}$ values in $a$ direction. In the 
right panel of Fig. \ref{rech_2} we compare the ladder systems with eight 
rungs [cluster (3) in Fig.~\ref{ebene}] and with four rungs. One observes 
large finite-size effects for the loss function, especially for small momentum 
transfer ${\bf q}$. Note that in particular, the larger clusters (2) and (3) 
from Fig.~\ref{ebene} (solid lines in Fig.~\ref{rech_1}) lead to additional 
peaks in the spectra. This clearly demonstrates that the original cluster (1) 
in Fig.~\ref{ebene}, which is formed of two ladders with four rungs, is not 
large enough to allow reliable conclusions. 

\begin{figure}
  \begin{center}
    \scalebox{0.59}{
      \includegraphics*[105,420][510,790]{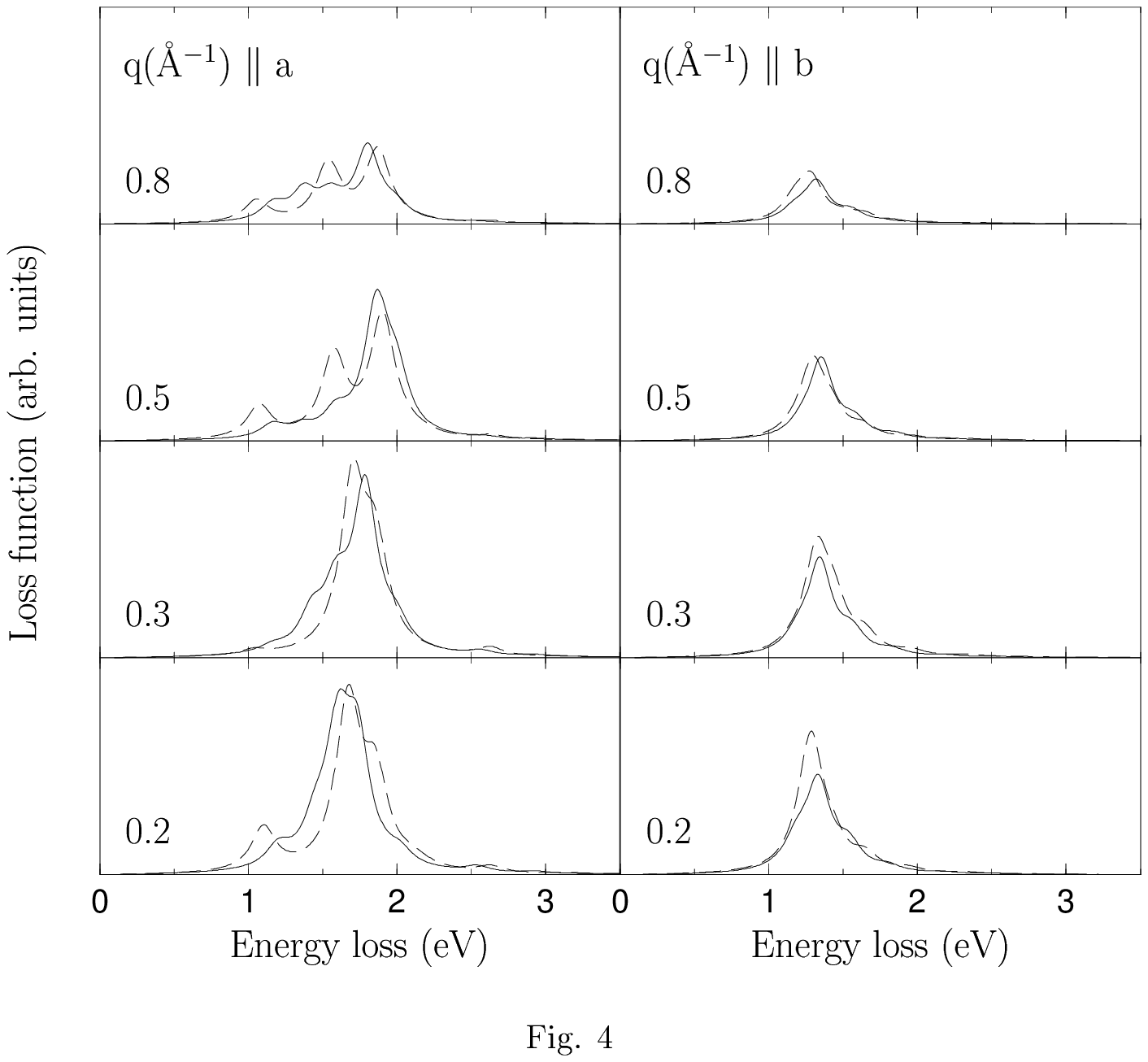}
    }
  \end{center}
  \caption{
    Loss function for open boundary conditions (full lines) and periodic
    boundary conditions in ${\bf q}$ direction (dashed lines); parameters as
    for Fig.~\ref{rech_1}.
  }
  \label{rech_3}
\end{figure}

To test if there are still finite size-effects for clusters (2) and (3), we 
next compare the spectra for closed and open boundary 
conditions in direction of momentum transfer. In the case of open boundary 
conditions one has to make sure that electrons on the edges of the cluster 
are still embedded in the local Coulomb potential that results from a zig-zag 
ordered state. For this purpose, sites on the edge of cluster (2) in 
Fig.~\ref{ebene} are assigned an additional on-site energy $V_{xy}$. Sites on 
the edges of cluster (3) that are not occupied in a zig-zag charge 
ordered state need an additional on-site energy $V_{b}$. As can be seen from 
Fig.~\ref{rech_3} (left panel) there are only small differences between 
periodic (in ${\bf q}$ direction) and open boundary conditions for momentum 
transfer parallel to $a$ direction, whereas one observes systematic 
discrepancies for small momentum transfer parallel to $b$ direction [right 
panel of Fig.~\ref{rech_3}]. For ${\bf q}<0.2$\AA$^{-1}$ there are large 
differences in both directions (not shown). We conclude that the convergence 
of the loss function is satisfactory for momentum transfer 
${\bf q}\ge 0.2$\AA$^{-1}$. In the next section we use clusters (2) and (3) of 
Fig.~\ref{ebene} with periodic boundary conditions parallel to ${\bf q}$ 
direction for the calculation of the electron-energy loss spectrum.


\section{Electron-energy loss spectrum}\label{Elect}

For the comparison with the experimental loss function\cite{Atzkern} of 
NaV$_{2}$O$_{5}$ we use the full $t$-$U$-$V$ model \eqref{hamilton}. The 
hopping parameters and on-site Hubbard interactions are taken from 
Ref.~\ref{Gros}. The values of the intersite Coulomb interactions 
$V_{a}=0.8$eV, $V_{b}=0.6$eV and $V_{xy}=0.9$eV have been adjusted to obtain 
correct peak positions of the loss function. 

\begin{figure}
  \begin{center}
    \scalebox{0.715}{
      \includegraphics*[135,300][470,800]{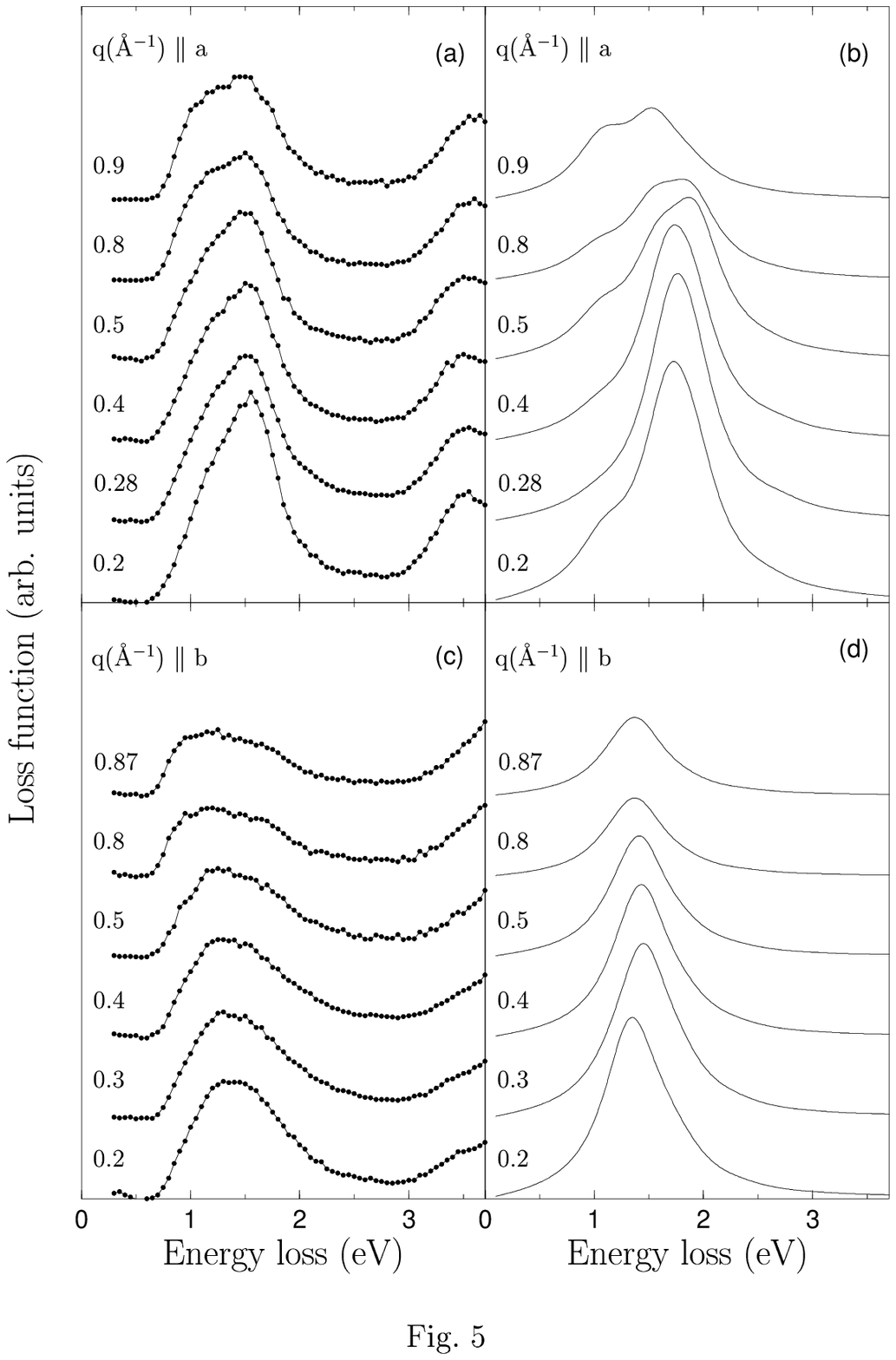}
    }
  \end{center}
  \caption{
    Comparison of experimental data for NaV$_{2}$O$_{5}$ (left), taken from
    Ref.~\ref{Atzkern}, and the calculated loss function plotted with an
    energy resolution of $0.3$eV (right). The hopping parameters and the
    on-site Hubbard interaction of the $t$-$U$-$V$ model are used from
    Ref.~\ref{Gros}, the intersite Coulomb interactions are
    $V_{a}=0.8$eV, $V_{b}=0.6$eV, and $V_{xy}=0.9$eV.
  }
  \label{vergleich}
\end{figure}

In Fig.~\ref{vergleich} the obtained loss function for the full model 
\eqref{hamilton} is compared to the experimental spectra taken from 
Ref.~\ref{Atzkern}. Very good agreement for the loss function with momentum 
transfer parallel to $a$ direction [see panel (a) and (b) of 
Fig.~\ref{vergleich}] is found. In particular, the increasing width of the 
experimentally observed structure at 1-2eV with increasing {\bf q} is 
reproduced. On the other hand, for momentum transfer parallel to $b$ direction 
[panel (d) of Fig.~\ref{vergleich}] the observed structure at $1.0$-$1.7$eV is 
not broad enough compared to the experiment [panel (c) of 
Fig.~\ref{vergleich}]. However, the agreement for momentum transfer parallel 
to $b$ direction is significantly better for the full $t$-$U$-$V$ model 
\eqref{hamilton} than for the $t$-$J$-$V$ model 
\eqref{ef_hamilton}.\cite{Atzkern} Note that for momentum transfer in $a$ 
direction the results for both models differ only slightly. This implies that 
there are different basic mechanisms for excitations with momentum transfer 
in $a$ or $b$ direction: doubly occupied sites are only important for momentum 
transfer parallel to $b$ direction.

Now let us discuss the character of these excitations in more detail. In the 
ground state all rungs of the ladders are found to be essentially singly 
occupied. Since the system is close to the charge order 
transition,\cite{Vojta1,Cuoco} one can illustrate the nature of the 
excitations using a zig-zag ordered ground state.

\begin{figure}
  \begin{center}
    \scalebox{0.75}{\includegraphics*[141,365][455,515]{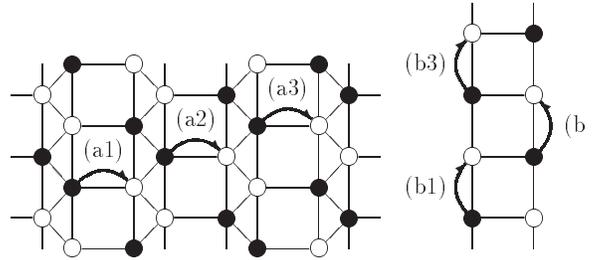}}
  \end{center}
  \caption{
    Illustration of the basic excitations with momentum transfer parallel to
    $a$ direction (left) and $b$ direction (right). In addition to the local
    transition on one rung [process (a1)], for momentum transfer parallel to
    $a$ direction one also observes collective transitions of two or three
    adjacent rungs on different ladders coupled by the interladder Coulomb
    interaction $V_{xy}$ [process (a1)+(a2) or (a1)+(a2)+(a3)]. The creation
    of one unoccupied and one doubly occupied rung [process (b1)] is the basic
    mechanism for excitations with momentum transfer parallel to $b$ direction.
    Electron movement can lead to an increased distance between unoccupied and
    doubly occupied rung [processes (b1)+(b2), (b1)+(b2)+(b3)].
  }
  \label{process}
\end{figure}

Disturbance of the charge ordering by electron hopping on a rung [process (a1) 
in Fig.~\ref{process}] is found to be the basic mechanism for excitations with 
momentum transfer parallel to $a$ direction. The excitation energy is 
dominated by $V_{b}$, and other rungs of the ladders are only involved via 
$V_{b}$ and $V_{xy}$. Note that process (a1) can also be interpreted as a 
transition from a bonding to an anti-bonding state of a singly occupied 
rung.\cite{Damascelli2000} In addition to this one electron transition, there 
are also collective transitions that involve two [process (a1)+(a2) in 
Fig.~\ref{process}] or three [process (a1)+(a2)+(a3) in Fig.~\ref{process}] 
adjacent rungs on different ladders. The collective nature of these 
transitions results from the interladder Coulomb interaction $V_{xy}$. The 
role of $V_{xy}$ can best be illustrated in the limit of small electron 
hopping (see Fig.~\ref{process}). In this case the excitation energies are 
$2V_{b}$, $(4V_{b}-V_{xy})$, and $(6V_{b}-2V_{xy})$ for the processes (a1), 
(a1)+(a2), and (a1)+(a2)+(a3).

The theoretical spectra for momentum transfer parallel to $a$ direction [see 
panel (b) of Fig.~\ref{vergleich}] mainly consist of three excitations at 
$1.2$eV, $1.6$eV, and $1.8$eV that contribute to the structure at $1$-$2$eV 
energy loss. With increasing momentum transfer ${\bf q}$ spectral weight 
shifts from the excitation at $1.6$eV to the one at $1.8$eV. The three 
excitations differ in the contribution of processes shown in 
Fig.~\ref{process}. The excitation at $1.2$eV is rather delocalized and 
consists mainly of multi-electron transitions [processes (a1)+(a2) or 
(a1)+(a2)+(a3) in Fig.~\ref{process}]. The other peaks result from more 
localized processes that are dominated by one electron transitions [process 
(a1) in Fig.~\ref{process}]. In addition, the transition (a1)+(a2) in 
Fig.~\ref{process} contributes to the excitation with energy loss $1.6$eV. 

The excitations that correspond to the structure at $1.0$-$1.7$eV for momentum 
transfer parallel to $b$ direction [see panel (d) of Fig.~\ref{vergleich}] can 
be interpreted as transitions to states with one unoccupied and one doubly 
occupied rung [process (b1) in Fig.~\ref{process}]. The excitations differ in 
the distance between unoccupied and doubly occupied rung, see processes 
(b1)+(b2) and (b1)+(b2)+(b3). These excitations are not influenced by the 
interladder Coulomb interaction $V_{xy}$, since the contribution of the 
$V_{xy}$ part of Hamiltonian \eqref{hamilton} is not changed by electron 
movement in one ladder of the zig-zag ordered system at quarter filling, and 
an unoccupied or a doubly occupied rung can be easily moved along the ladder.  
Since we can only observe rather localized excitations by direct 
diagonalization of small clusters there is a lack of spectral weight in our 
theoretical spectra [panel (d) of Fig.~\ref{vergleich}] compared to the 
experimental data [panel (c) of Fig.~\ref{vergleich}].


\section{Optical conductivity}\label{optic_cond}

\begin{figure}
  \begin{center}
    \scalebox{0.59}{
      \includegraphics*[107,420][515,790]{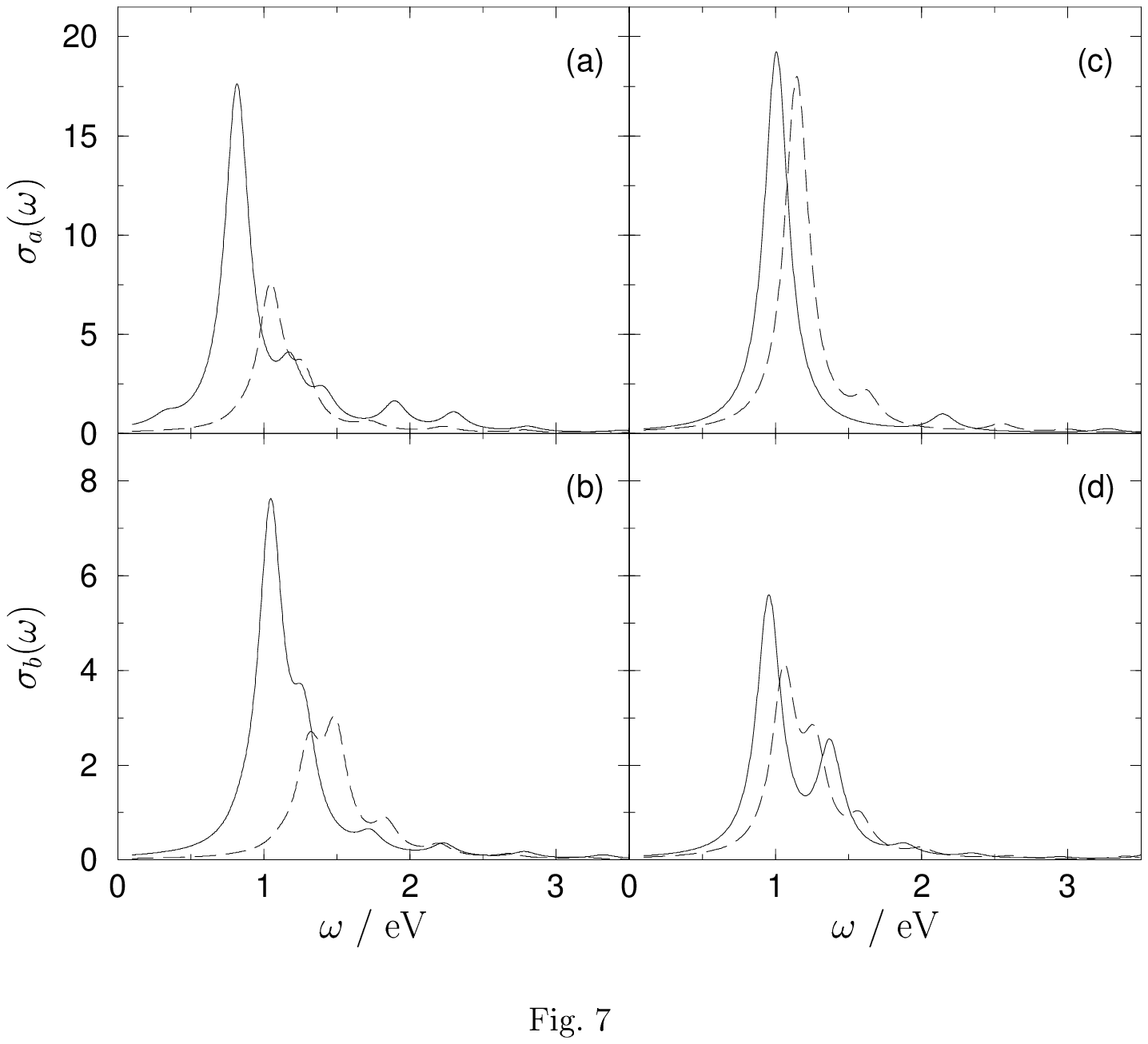}
  }
  \end{center}
  \caption{
    Optical conductivity of cluster (1) in Fig.~\ref{ebene} [panels (a) and
    (b), parameters of the $t$-$J$-$V$  model are used from Ref.~\ref{Cuoco}],
    cluster (2) and (3) [panels (c) and (d), parameters of the $t$-$U$-$V$
    model as for Fig.~\ref{vergleich}] plotted with an energy resolution of
    $0.1$eV for periodic boundary conditions (full lines) and open boundary
    conditions (dashed lines).
  }
  \label{optik}
\end{figure}

Before we discuss our results for the optical conductivity, we briefly study 
their finite-size effects. A good agreement between calculated and 
experimental optical conductivity was obtained from a finite temperature 
Lanczos method.\cite{Cuoco} In Ref.~\ref{Cuoco} cluster (1) in 
Fig.~\ref{ebene} with periodic boundary conditions was used, and the 
parameters of the $t$-$J$-$V$ model \eqref{ef_hamilton} were chosen as 
$t_{a}=0.4$eV, $t_{b}=0.2$eV, $t_{xy}=0.15$eV, $U=4$eV, $V_{a}=V_{b}=0.8$eV, 
and $V_{xy}=0.9$eV. However, for this system we obtain large finite-size 
effects as can be seen from a comparison of the results for periodic [full 
lines in panel (a) and (b) of Fig.~\ref{optik}] and open boundary conditions 
[dashed lines in panel (a) and (b) of Fig.~\ref{optik}]. Consequently, the 
cluster formed of two ladders with four rungs is not large enough to obtain 
reliable results for the optical conductivity. To overcome this problem we 
again use clusters (2) and (3) of Fig.~\ref{ebene} for the calculation of 
$\sigma_{a}$ and $\sigma_{b}$. (In Sec.~\ref{finite} we have shown that a 
reduction of the cluster perpendicular to the momentum transfer affects the 
loss function only weakly if the interladder hopping $t_{xy}$ is small. This 
statement remains valid for the calculation of the optical conductivity if the 
direction of the momentum transfer is replaced by the direction of the 
electric field.)

For the comparison with the experimental optical conductivity\cite{Damascelli} 
of NaV$_{2}$O$_{5}$ we use the full $t$-$U$-$V$ model \eqref{hamilton}, and 
the same parameters as for the calculation of the loss function in 
Sec.~\ref{Elect}. In panels (c) and (d) of Fig.~\ref{optik} the results for 
$\sigma_{a}$ and $\sigma_{b}$ using clusters (2) and (3) of Fig.~\ref{ebene} 
are shown. As a result of finite-size effects discussed in Sec.~\ref{finite} 
we obtain only a qualitative agreement of the calculated optical conductivity 
and the experimental data of Ref.~\ref{Damascelli}. In analogy to the loss 
function, there is a lack of spectral weight for $\sigma_{b}$, and the ratio 
$r=I_{a}/I_{b}=3.3$ of total $a$ and $b$ intensities is larger than the 
experimental value of $2.2$. Note, however, that the differences of the 
optical conductivity for periodic (full lines) and open boundary conditions 
(dashed lines) using clusters (2) and (3) of Fig.~\ref{ebene} are much smaller 
than the differences for the cluster (1). In contrast to 
Ref.~\onlinecite{Cuoco} we have used a small interladder hopping $t_{xy}$ to 
describe the optical conductivity. Consequently, we conclude from the 
comparison with the experiment that interladder hopping is of minor importance 
for a realistic description of charge excitations in NaV$_{2}$O$_{5}$. The 
same was found in a previous theoretical study\cite{Nishimoto1} of the optical 
conductivity where a $t$-$J$-$V$ model was investigated which was slightly 
modified by an additional symmetry-breaking on-site energy.


\section{Conclusions}\label{conclusion}

Summing up, we have calculated the EELS spectrum and the optical conductivity 
for the quarter-filled ladder compound $\alpha'$-NaV$_{2}$O$_{5}$ by exact 
diagonalization of small systems using the standard Lanczos algorithm. Our 
analysis shows that finite-size effects for these calculations are very 
important, in particular for the loss function with small momentum transfer 
and for the optical conductivity. We minimize these finite-size effects by 
using clusters of different shape depending on the direction of momentum 
transfer and the direction of the electric field respectively. The results for 
the loss function are in good agreement with experimental spectra. We also 
obtain a qualitative description of the optical conductivity. The comparison 
with the experiment confirms that a large value of the interladder hopping is 
not needed for a realistic description of charge excitations in 
NaV$_{2}$O$_{5}$. We find that the basic mechanism for excitations strongly 
depends on the direction of momentum transfer. For momentum transfer parallel 
to $a$ direction three different excitations are observed. They differ in 
their degree of delocalization. The collective character of the delocalized 
processes results from the interladder Coulomb interaction. The excitations 
for momentum transfer parallel $b$ direction can be characterized by the 
creation of an unoccupied and a doubly occupied rung.


\section*{Acknowledgements}

Discussions with S. Atzkern, J. Fink, M. S. Golden, R. E. Hetzel, J. Richter, 
and M. Vojta are gratefully acknowledged. This work was supported by DFG 
through the research program of the GK 85, Dresden. The calculations were 
performed on the Origin 2000 at Technische Universit\"{a}t Dresden.


\end{multicols}
\widetext


\begin{references}
  \bibitem{Weiden}M. Weiden, R. Hauptmann, C. Geibel, F. Steglich, M. Fischer, 
           P. Lemmens, and G. G\"{u}ntherodt, Z. Phys. B {\bf 103}, 1 (1997).
  \bibitem{Isobe}M. Isobe and Y. Ueda, J. Phys. Soc. Jpn. {\bf 65}, 1178 
           (1996).
  \bibitem{Fujii}Y. Fujii, H. Nakao, T. Yoshihama, M. Nishi, K. Nakajima, 
           K. Kakurai, M. Isobe, Y. Ueda, and H. Sawa, J. Phys. Soc. Jpn. 
           {\bf 66}, 326 (1997).
  \bibitem{Capry}A. Capry and J. Galy, Acta Cryst. {\bf 31}, 1481 (1975).
  \bibitem{Schnering}H. G. Schering, Y. Grin, M. Kaupp, M. Sommer, 
           R. K. Kremer, O. Jepsen, T. Chatterji, and M. Weiden, 
           Z. Kristallogr. {\bf 213}, 246 (1998).
  \bibitem{Gros}H. Smolinski, C. Gros, W. Weber, U. Peuchert, G. Roth, 
           M. Weiden, and C. Geibel, Phys. Rev. Lett. {\bf 80}, 5164 
           (1998).\label{Gros}
  \bibitem{Horsch}P. Horsch and F. Mack, Eur. Phys. J. B {\bf 5}, 367 (1998).
           \label{Horsch}
  \bibitem{Koeppen}M. K\"{o}ppen, D. Pankert, R. Hauptmann, M. Lang, M. Weiden,
           C. Geibel, and F. Steglich, Phys. Rev. B {\bf 57}, 8466 (1998).
  \bibitem{Ohama}T. Ohama, H. Yasuoka, M. Isobe, and Y. Ueda, Phys. Rev. B 
           {\bf 59}, 3299 (1999).
  \bibitem{Seo}H. Seo and H. Fukuyama, J. Phys. Soc. Jpn. {\bf 67}, 2602 
           (1998).
  \bibitem{Mostovoy}M. V. Mostovoy and D. I. Khomskii, Solid State Commun. 
           {\bf 113}, 159 (2000).
  \bibitem{Thalmeier}P. Thalmeier and P. Fulde, Europhys. Lett. {\bf 44}, 242 
           (1998).
  \bibitem{Fagot}Y. Fagot-Revurat, M. Mehring, and R. K. Kremer, Phys. Rev. 
           Lett. {\bf 84}, 4176 (2000).
  \bibitem{Luedecke}J. L\"{u}decke, A. Jobst, S. van Smaalen, E. Morr\'{e}, C. 
           Geibel, and H.-G. Krane, Phys. Rev. Lett. {\bf 82}, 3633 (1999).
  \bibitem{Smaalen}S. van Smaalen and J. L\"{u}decke, Europhys. Lett. {\bf 49},
           250 (2000).
  \bibitem{Boer}J. L. deBoer, A. Meetsma, J. Baas, and T. T. M. Palstra, Phys. 
           Rev. Lett. {\bf 84}, 3962 (2000).
  \bibitem{Ohama2}T. Ohama, A. Goto, T. Shimizu, E. Ninomiya, H. Sawa, M. 
           Isobe, and Y. Ueda, preprint (cond-mat/0003141).
  \bibitem{Bernert}A. Bernert, T. Chatterji, P. Thalmeier, and P. Fulde, 
    preprint (cond-mat/0012327).
  \bibitem{Fulde}P. Fulde, Ann. Physik {\bf 6}, 178 (1997).
  \bibitem{Vojta1}M. Vojta, R. E. Hetzel, and R. M. Noack, Phys. Rev. B
           {\bf 60}, R8417 (1999).
  \bibitem{Vojta2}M. Vojta, A. H\"{ubsch}, and R. M. Noack, Phys. Rev. B 
           {\bf 63}, 045105 (2001).
  \bibitem{Atzkern}S. Atzkern, M. Knupfer, M. S. Golden, J. Fink, 
           A. N. Yaresko, V. N. Antonov, A. H\"{u}bsch, C. Waidacher, 
           K. W. Becker, W. von der Linden, G. Obermeier, and S. Horn, Phys. 
           Rev. B {\bf 63}, 165113 (2001).\label{Atzkern}
  \bibitem{Nishimoto1}S. Nishimoto and Y. Ohta, J. Phys. Soc. Jpn. {\bf 67}, 
           3679 (1998).
  \bibitem{Nishimoto2}S. Nishimoto and Y. Ohta, J. Phys. Soc. Jpn. {\bf 67},
           4010 (1998).
  \bibitem{Schnatterly}S. E. Schnatterly, Solid State Phys. {\bf 34}, 275
           (1977).
  \bibitem{Long}V. C. Long, Z. Zhu, J. L. Musfeldt, X. Wei, H.-J. Koo, 
           M.-H. Whangbo, J. Jegoudez, and A. Revcolevshi, Phys. Rev. B 
           {\bf 60}, 15721 (1999).
  \bibitem{Presura}C. Presura, D. van der Marel, A. Damascelli, and R. K. 
           Kremer, Phys. Rev. B {\bf 61}, 15762 (2000).
  \bibitem{Yosihama}T. Yosihama, M. Nishi, K. Nakajima, K. Kakurai, Y. Fuji, 
           M. Isobe, C. Kagami, and Y. Ueda, J. Phys. Soc. Jpn. {\bf 67},
           744 (1998).
  \bibitem{Cuoco}M. Cuoco, P. Horsch, and F. Mack, Phys. Rev. B {\bf 60},
           R8438 (1999).\label{Cuoco}
  \bibitem{Lin}For example, see H. Q. Lin and J. E. Gubernatis, Computers in
           Physics {\bf 7}, 400 (1993), and references therein.
  \bibitem{Jaklic}J. Jakli\v{c} and P. Prelov\v{s}ek, Phys. Rev. B {\bf 49}, 
          5065 (1994).
  \bibitem{Damascelli}A. Damascelli, D. van der Marel, M. Gr\"{u}ninger, 
           C. Presura, T. T. M. Palstra, J. Jegoudez, and A. Revcolevschi, 
           Phys. Rev. Lett. {\bf 81}, 918 (1998).\label{Damascelli}
  \bibitem{Damascelli2000}A. Damascelli, C. Presura, D. van der Marel, 
           J. Jegoudez, and A. Revcolevschi, Phys. Rev. B {\bf 61}, 2535 
           (2000).


\end{references}
\end{document}